\documentclass[%
reprint,
amsmath,amssymb,
aps,
]{revtex4-1}
\usepackage{graphicx}
\usepackage{dcolumn}
\usepackage{bm}
\usepackage{subfigure}
\usepackage{soul}
\usepackage{xcolor, soul}
\sethlcolor{green}
\sethlcolor{red}

\usepackage{subfigure}
\usepackage{graphicx,epsfig,epstopdf}
\usepackage{amsmath}
\usepackage{color}
\usepackage{gensymb}

\begin{document}
	
	\preprint{APS/123-QED}
	
	\title{Spatial Wave Control Using a Self-biased Nonlinear Metasurface at Microwave Frequencies}
	
	
	\author{Mehdi Kiani$^1$} 
	\author{Ali Momeni$^2$}
	
	\author{Majid Tayarani$^1$}
	\email{m$ _ $tayarani@iust.ac.ir}
	
	\author{Can Ding$^3$}%
	\affiliation{%
		$^1$School of Electrical Engineering, Iran University of Science and Technology, Tehran, Iran.\\
		$^2$Applied Electromagnetic Laboratory, School of Electrical Engineering, Iran University of Science and Technology, Tehran, Iran\\
		$^3$Global Big Data Technologies Centre,University of Technology Sydney, Ultimo, NSW 2007, Australia
	}	
	\begin{abstract}
		Recently, investigation of metasurface has been extended to wave control through exploiting nonlinearity. Among all of the ways to achieve tunable metasurfaces with multiplexed performances, nonlinearity is one of the promising choices.
		Although several proposals have been reported to obtain  nonlinear architectures at visible frequencies, the area of incorporating nonlinearity in form of passive-designing at microwave metasurfaces is open for investigation.
		In this paper, a passive wideband nonlinear metasurface is manifested, which is composed of embedded L$-$shape and $\Gamma-$shape meta-atoms with PIN-diode elements. The proposed self-biased nonlinear metasurface has two operational states: at low power intensities, it acts as Quarter Wave Plate (QWP) in the frequency range from 13.24 GHz to 16.38 GHz with an Axial Ratio (AR) of over 21.2\%. In contrast, at high power intensities, by using the  polarization conversion property of the proposed PIN-diode based meta-atoms, the metasurface can act as a digital metasurface. It means that by arranging the meta-atoms with a certain coding pattern, the metasurface can manipulate the scattered beams and synthesize well-known patterns such as Diffusion-like pattern at an ultra-wide frequency range from 8.12 GHz to 19.27 GHz (BW=81.4\%). Full-wave and nonlinear simulations are carried out to justify the performance of the wideband nonlinear metasurface. We expect the proposed self-biased nonlinear metasurface at microwave frequencies reveals excellent opportunities to design limiter metasurfaces and compact reconfigurable imaging systems.
	\end{abstract}
	
	\maketitle
	
	\begin{figure*}[t]
		\centering
		\includegraphics[height=3.5in]{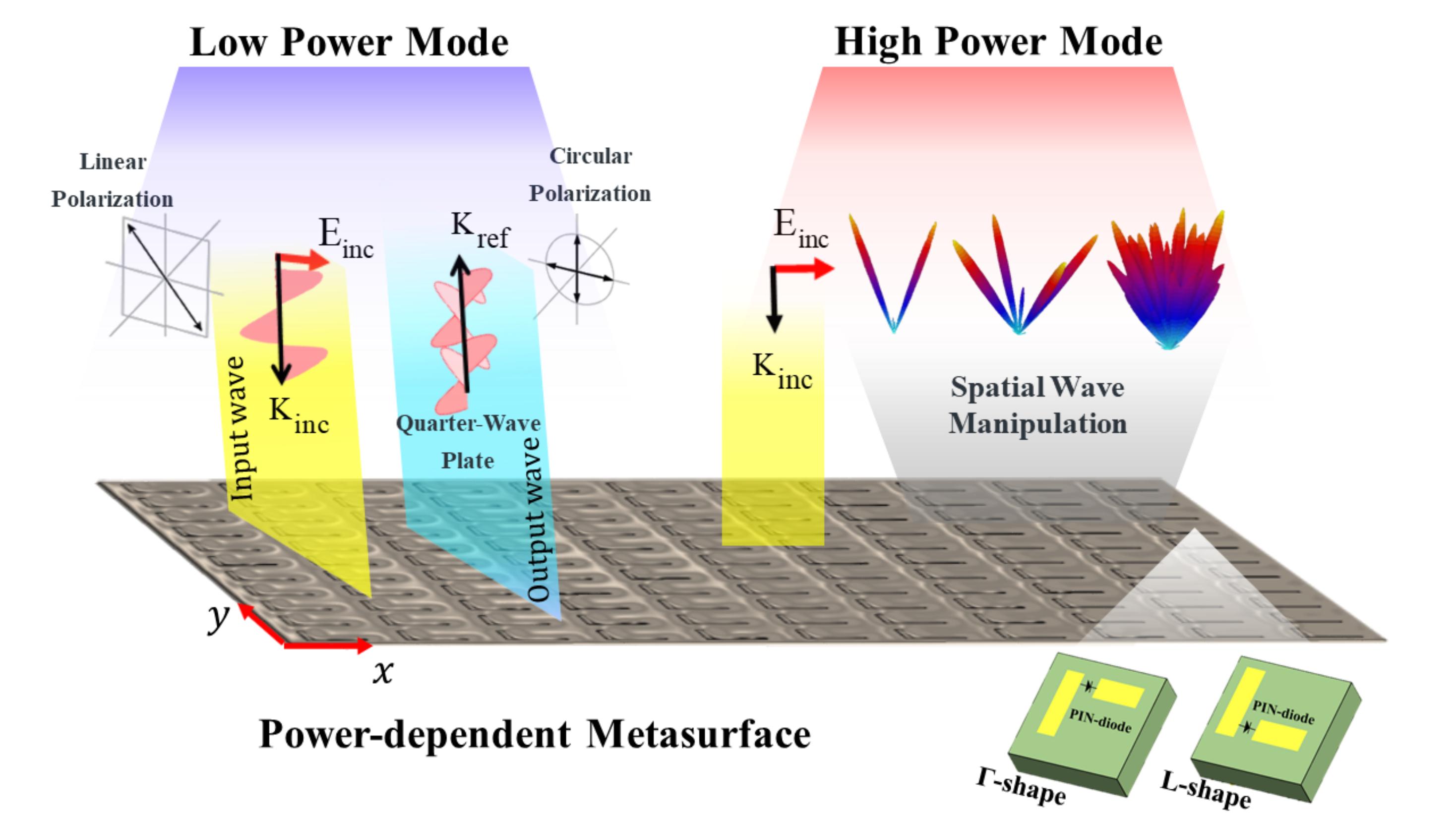}
		\caption{\label{fig:epsart} {The perspective view of the nonlinear metasurface. When a weak tilted linear-polarized wave is incident to the metasurface, it acts as a quarter-wave plate. Whenever a high power wave is illuminated the metasurface, the operational status changes to a 1-bit digital metasurface.  
		}}
	\end{figure*}
	\section{Introduction}
	
	At the beginning of the century, remarkable attention was absorbed to metasurfaces as the sub-wavelength structures owing to their electromagnetic features, besides being low loss and easy to chip. Metasurfaces have created a broad area to wave control from microwave to optical frequencies. Meanwhile, theoretical developments like the emergence of Generalized Snell’s law of reflection and refraction, paved this way. The result of the rapid growth of metasurfaces to wave control has witnessed and found a special place in achieving negative refraction \cite{shelby2001experimental,ziolkowski2003design}, engineered antennas \cite{ziolkowski2010metamaterial,movahhedi2020dual,movahhedi2019multibeam}, beam scanning \cite{moeini2019wide,moeini2019collimating}, invisibility cloak \cite{pendry2006controlling,li2008hiding}, absorbers \cite{rahmanzadeh2018multilayer}, polarizers \cite{zhao2011manipulating}, dispersion engineering \cite{feng2012engineering, guo2015dispersion}, and advanced analog computing \cite{babaee2020parallel, momeni2019generalized,abdolali2019parallel, rajabalipanah2020space, momeni2019tunable}. Demands for any type of flexible structures made scientists of this field use the materials like graphene \cite{rouhi2020wideband,fallahi2012design,cheng2015dynamically}, vanadium dioxide \cite{liu2016hybrid,cueff2015dynamic,hashemi2016electronically} and semiconductors \cite{ratni2018active,li2017electromagnetic} as the phase change materials. In addition to the use of unique materials to obtain metasurfaces with different functionalities, nonlinearity can also help to achieve controllable metasurfaces. In recent years, especially at visible frequencies by using diverse nonlinear materials, several numbers of studies have incorporated nonlinearity into the metasurface design for many applications such as polarization rotation \cite{rose2013circular}, harmonic generation \cite{rose2011controlling}, frequency multiplexing \cite{ye2016spin}, and so forth. More recently, at microwave frequencies, with the help of semiconductor lumped elements like PIN-diodes and varactors which embedded into the unit cells, active power-dependent metasurfaces have been presented. Varactors at different powers can display different behaviors, and the power levels verify the capacitance of them in each time \cite{ratni2018active}. On the other side, PIN-diode at low power intensities acts as an open circuit or more precisely like a small capacitance, and at high power intensities when the power intensity is more than a specified threshold, it operates as a short circuit or small resistance \cite{zhao2019power}. Sievenpiper’s group in 2013  by introducing a high power surface wave circuit-based metasurface absorber\cite{wakatsuchi2013circuit} opened new doors to reach reconfigurable metasurfaces through the use of the nonlinear property at microwave frequencies. In 2017 \cite{zhao2017power}, a frequency selective surface has introduced to provide passband for high power intensities which could spectrally manipulate incoming wave; this study similar to Sievenpiper’s work profits from nonlinearity characteristic at microwave frequencies, using integrating PIN diodes into the different particles of the metasurface. Digital coding metasurfaces have created an opportunity to spatially re-distribute scattered waves via two distinct coding meta-atoms with opposite reflection/transmission phases (e.g., 0$^\circ$ and 180$^\circ$) as the $ "0" $ and $ "1" $ digital bits form the metasurface \cite{cui2014coding,rouhi2018real,rouhi2019multi}. 
	More recently, Cui’s group proposed two intensity-dependent metasurfaces whose functionality of them have been determined by the power intensity of the incoming wave \cite{luo2019intensity,luo2019digital}. Organizing coding particles with the help of varactor elements into the metasurface give them an opportunity to control scattered waves spectrally and spatially. In fact, “digital coding” and "nonlinearity" has emerged as a fascinating paradigm due not only to their capabilities of digitally manipulating EM waves, but also to synthesizing a metasurface with multiplexed performances. Moreover, recently a self-biased dual-band nonlinear metasurface has been proposed by authors \cite{kiani2020self} which at low power functions like an EM-mirror, while at high power levels shows two separate camouflage behaviors at C- and X-band.  However, similar to the other previous works, they achieve restricted functionalities at low power state, and moreover the operating band of the digital metasurface was limited to a single narrow frequency region that is unattractive for power limiting applications.
	
	In this paper, an ultrawideband self-biased nonlinear metasurface is presented, comprising of two types of nonlinear power-multiplexed L$-$shape and $\Gamma-$shape meta-atoms. At low-power intensities, the metasurface acts as a QWP that means it converts  Linear Polarized (LP) waves to Circular Polarized (CP) waves at low-power state. By contrast, At high power intensities,  the metasurface acts as a Digital Spatial Wave Manipulator Metasurface (DSWMM) by reflecting the LP incoming wave in cross-polarization and owing to deploying two nonlinear anti-phase meta-atoms. Indeed, the integrated PIN-diodes in the meta-atoms directly interact with the incoming wave and all its functionalities is solely dictated by the power intensity of the incident wave. Hence, our metasurface synthesizing is fully passive, it does not require any additional active biasing network, and, moreover, its performances are wideband enough for a wide diversity of practical applications demanding protection of sensitive devices from undesired high power signals. The numerical simulations verify the performance of the proposed bi-functional nonlinear coding metasurface.
	\section{Theory of analysis and design}
	\begin{figure*}[t]
		\centering
		\includegraphics[height=3in]{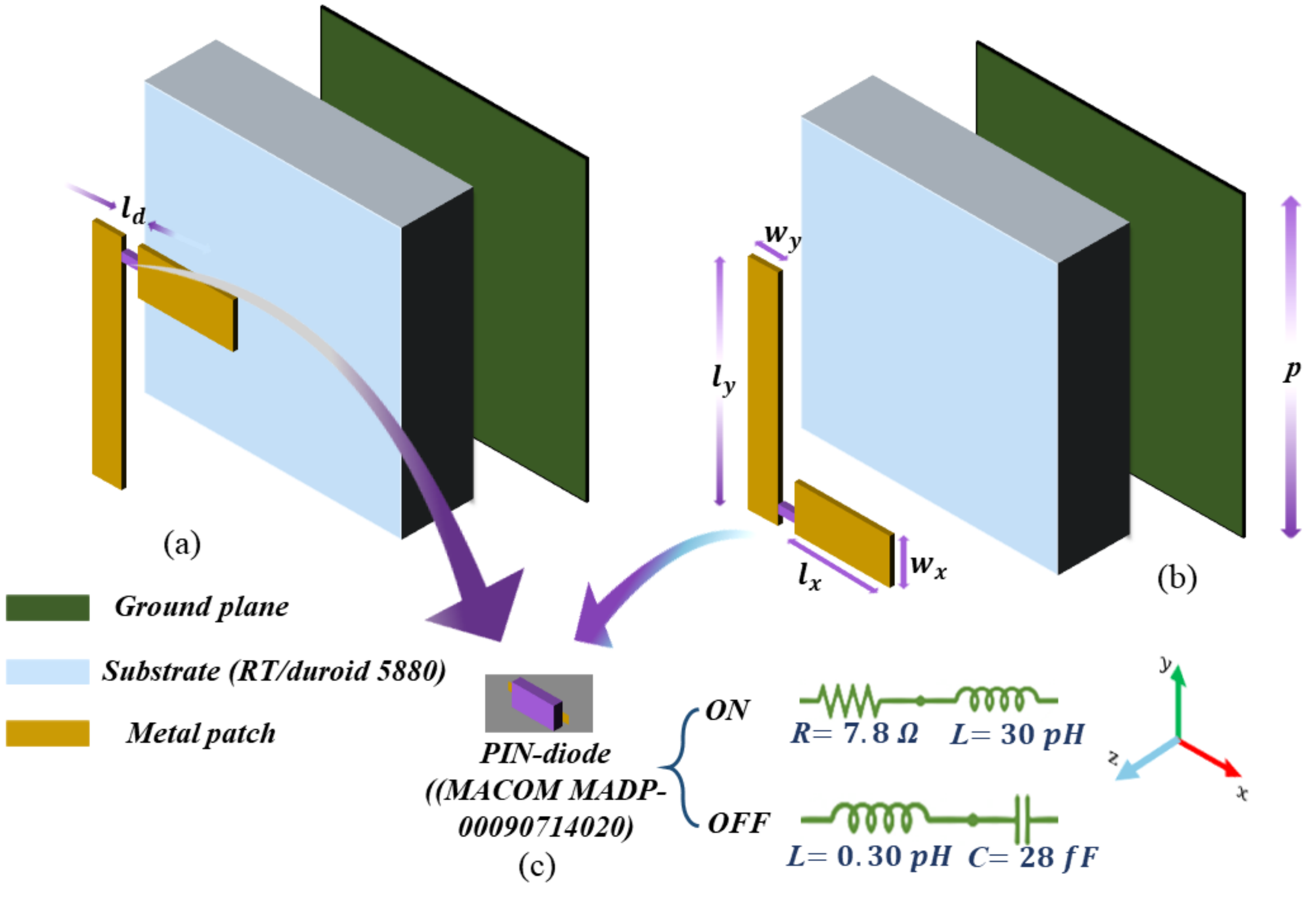}
		\caption{\label{fig:epsart} {(a), (b) The $\Gamma-$shape and L$-$shape meta$ - $atoms as the basic units of the metasurface. (c) circuit model of the embedded PIN-diodes for on- and off-states. }}
	\end{figure*}
	\subsection{Principle and design}
	
	\textcolor{blue}{Fig. 1} shows a perspective view of the nonlinear metasurface, which is composed of L$-$shape and $\Gamma-$shape  PIN-diode integrated meta$ - $atoms. The overall geometry of the proposed metasurface is described in \textcolor{blue}{Fig. 2(a, b)} consisting of three layers: 1) the L-shape metallic part as the top layer, 2) a 3.683 mm RT/duroid 5880 substrate, and 3) a copper ($\sigma=5.96 \times 10^7 $ S/m) ground plane as the bottom layer to zero transmission. The PIN-diode can be circuitously modelled as an $ RL $-circuit ($L$=30 pH, $R$=7.8 \ohm) in the on-state while it can be represented by an $ LC $-circuit ($L$=30 pH, $C$=0.028 pF) in the off mode \cite{yang2016programmable}. The common structural parameters of the proposed meta-atom are $ p $=10.6 mm, $ w_x $=2.1 mm, $ w_y$=1.5 mm, $ l_d $=0.6 mm, $l_y$=6.6 mm, and $ l_x $=4.2 mm. By changing the power intensity of the incoming wave, the electromagnetic responses of the proposed metasurface can be tailored, leading to a QWP at low power intensities and a DSWMM at high power intensities. At low power intensities, when PIN-diodes are in off-state and electric field vector slanted 45\degree relative to both $x-$ and $y-$directions the metasurface acts as a QWP. However, upon illuminating by high-power radiations, the functionality of the proposed device is changed; the metasurface can reflect LP waves to cross-polarized waves such that the L$-$shape and $\Gamma-$shape meta$ - $atoms mimic the digital states "0" and "1", realizing specific coding pattern over the surface.  The operational principle of the aforementioned metasurface is based on the circuit model of the PIN-diode which is changed from an $LC$-circuit to $RL$-circuit through increasing the power level of the illuminating wave. It is clear that the voltage levels induced on both ends of the embedded diode
	powerfully rely on the local field intensities stimulated by different incident power levels. Therefore, Depending on the induced voltage levels of PIN-diodes, we will see different electromagnetic responses. At low power illuminations, the peak of the induced voltage of PIN-diodes is not high enough to turn them on, therefore, they are modelled as open circuits (off-state) or more precisely as $LC$-circuits. Conversely, at high power illuminations, when the induced voltage of PIN-diodes is more than the threshold, they act as short circuits (on-state) or more accurately as $RL$-circuits \cite{coldren2012diode}(see \textcolor{blue}{Fig. 2(c)}). In spite of the previous metasurface demonstrations \cite{luo2019digital,luo2019intensity,yoo2014active}, the proposed nonlinear metasurface controls the off- and on-states of the PIN-diode without any auxiliary biasing network so that the operational status of the diodes is dictated by the power intensity of the incident wave. During the time variations of the high power AC illuminating signal, majority carriers (holes of P-type and electrons of N-type semiconductors) are stored in the intrinsic layer of PIN-diodes. At low frequencies, because recombination time of minority carriers (holes of N-type and electrons of P-type semiconductors) as the switching speed of the diode from on-to-off are in the order of time-domain signal period (for the exploited PIN-diode it is 2 nS), the stored charges can be detached from the intrinsic layer; therefore, the diode can switch to off mode. Conversely, at high frequencies, the recombination time of the minority carriers has a considerable value against the period of the input signal (if we consider the frequency range of simulations between 5 to 20 GHz, the corresponding time periodicity range will be T=50 ps to 200 pS). Consequently, there is not sufficient time to switch off the PIN-diode, and at high power illuminations, the PIN-diode will work stably in the on-state \cite{coldren2012diode}.
	\begin{figure*}%
		\centering
		\subfigure[]{%
			\label{fig:21}%
			\includegraphics[height=2in]{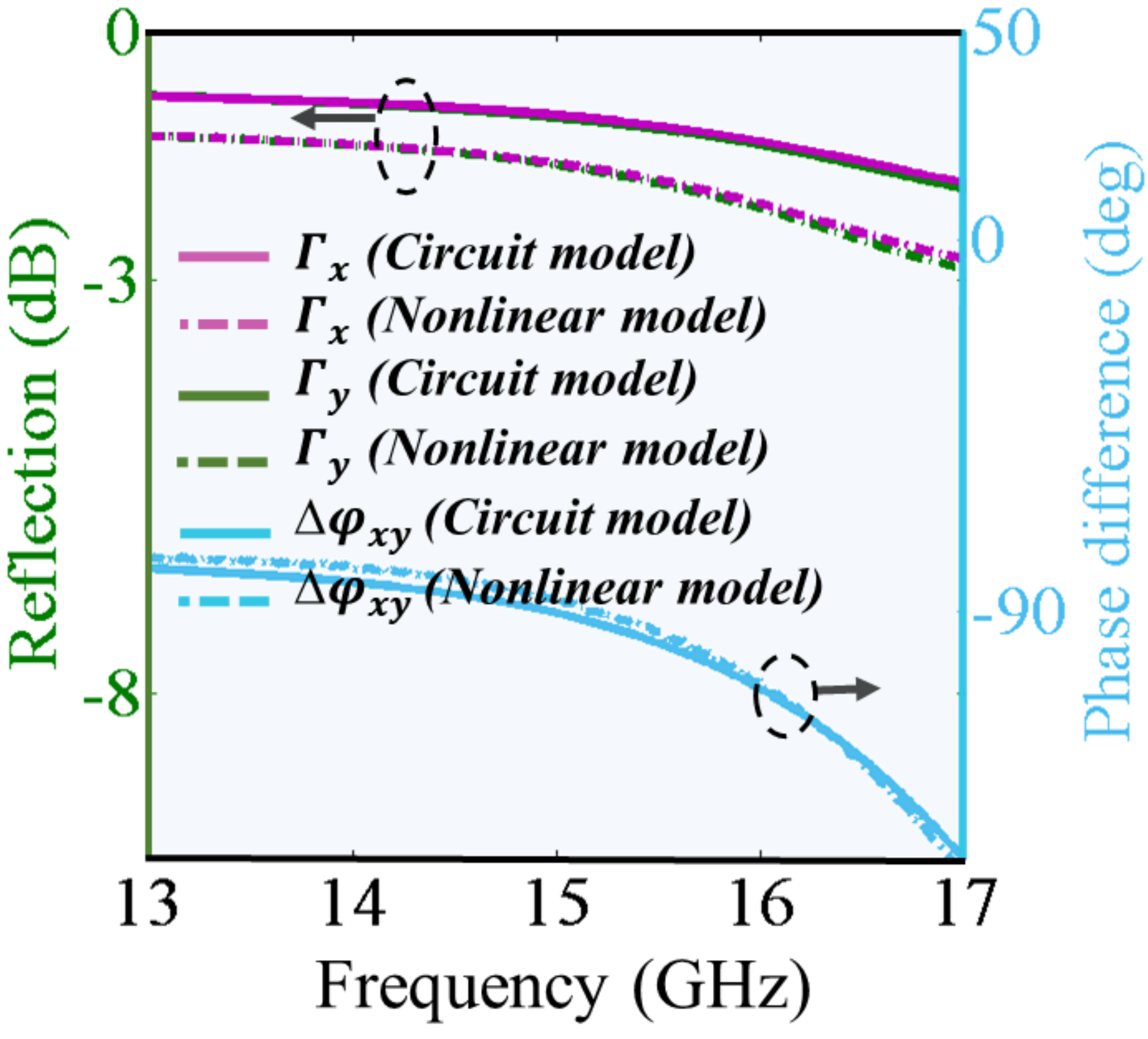}}
		\subfigure[][]{%
			\label{fig:22}%
			\includegraphics[height=2.02in]{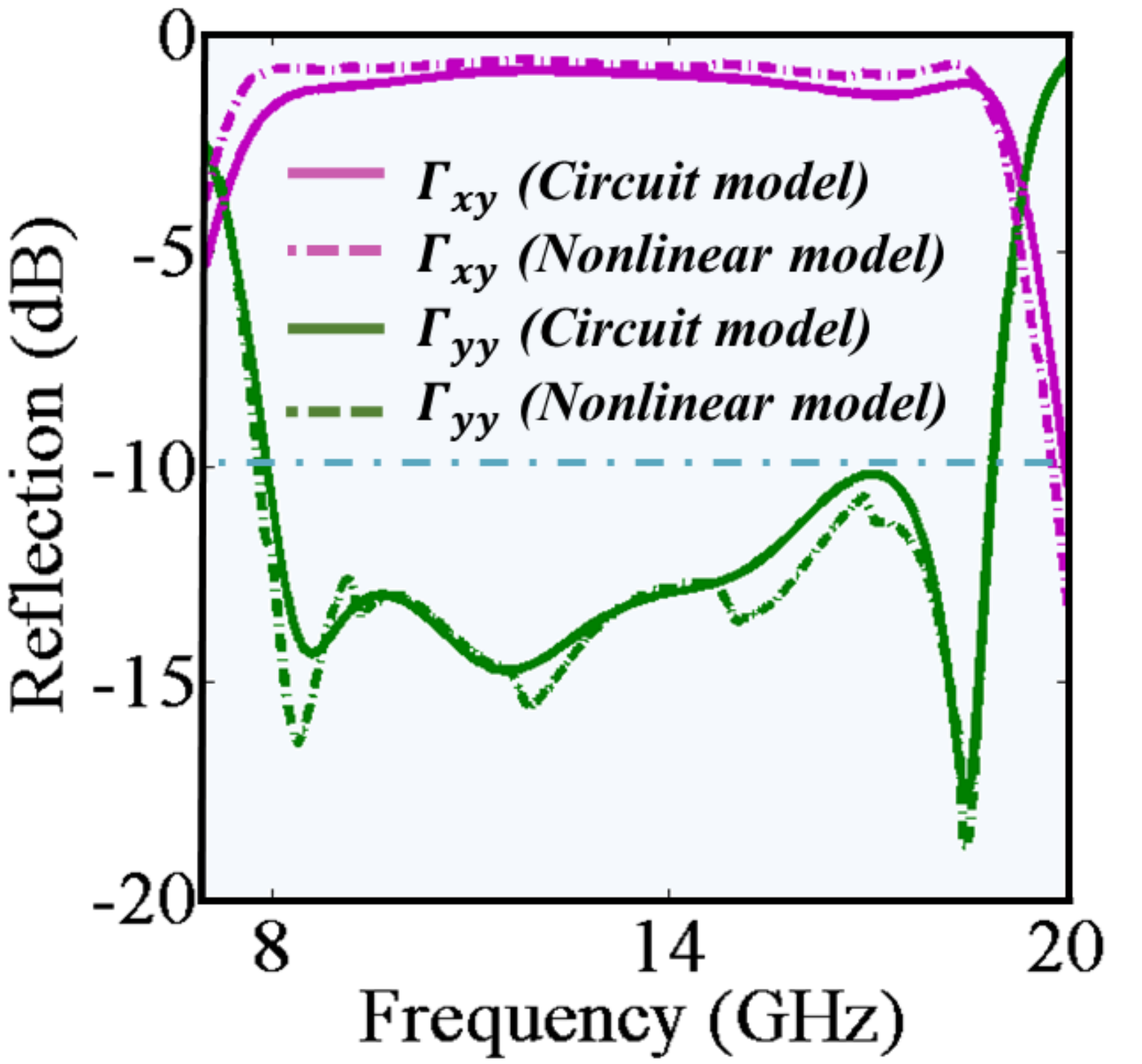}}%
		\qquad
		\subfigure[][]{%
			\label{fig:23}%
			\includegraphics[height=2.02in]{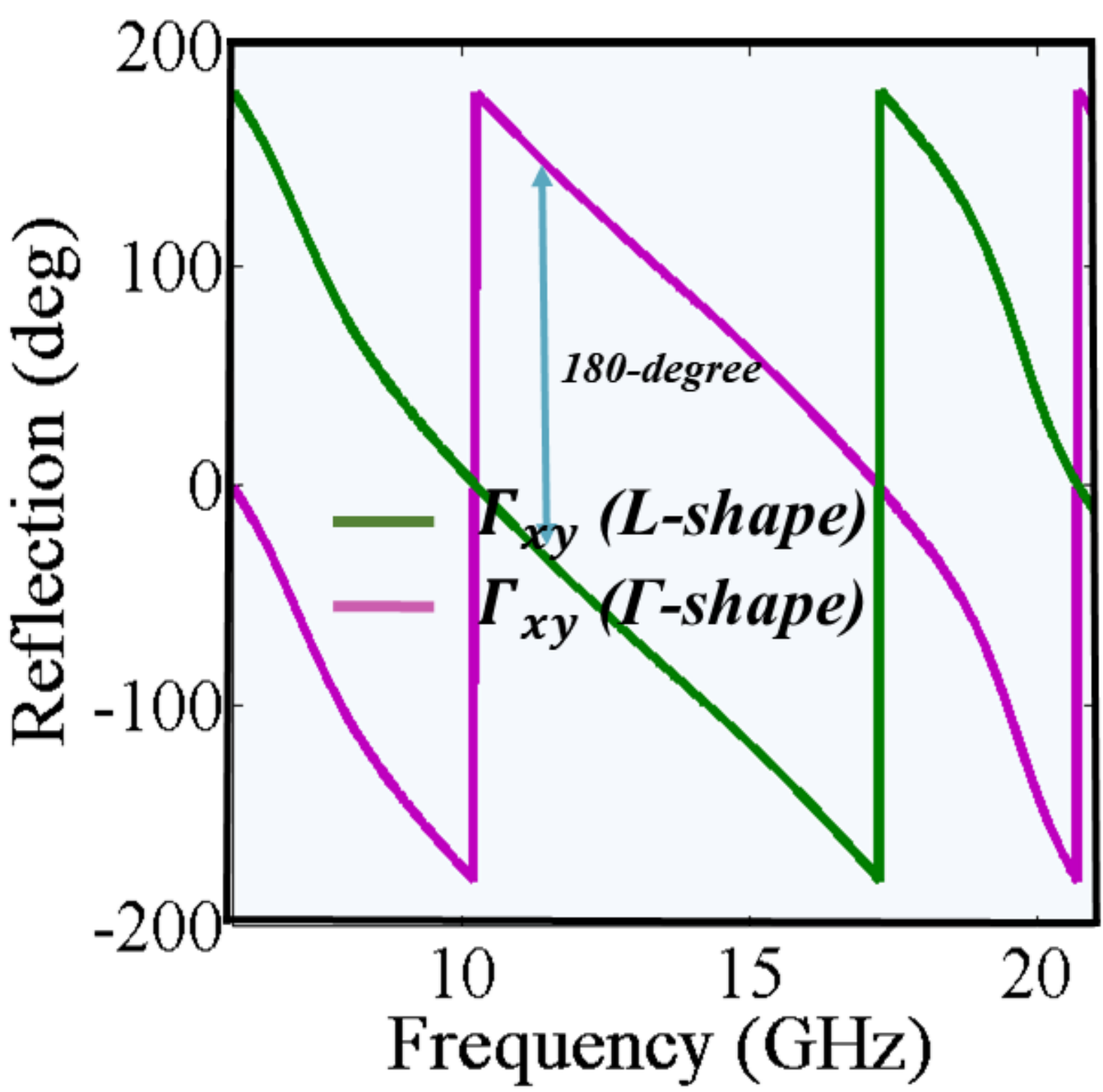}}%
		\caption{(a) The simulated reflection coefficients $\Gamma_x$ and $\Gamma_y$ and their phase difference $\Delta \varphi_{yx}$ of the QWP with simplified $ LC $ circuit and nonlinear (SPICE) model.(b)  The simulated reflection coefficients $\Gamma_{xy}$ and $\Gamma_{yy}$ of the DSWMM with simplified $ RL $-model and nonlinear (SPICE) model(c) The simulated phase difference between L$ - $shape and $ \Gamma $-shape meta$ - $atoms at high power intensities.}
		\label{fig:4}
	\end{figure*}
	\subsection{Frequency-domain analysis}
	To study the performance of the proposed metasurface. we consider that it is illuminated by a travelling wave along the $z-$direction; the reflected wave can be described as the sum of two LP components:
	\begin{equation}
	{\vec{E^r}} = {E_0}({\Gamma_x}\vec{x} + {\Gamma_y}\vec{y})
	\end{equation}
	in which, $E_0$ is the magnitude of the incident wave in both $x$- and $y$-directions. Also,  $\Gamma_x = |\Gamma_x|e^{\varphi_x}$ and $\Gamma_y= |\Gamma_y|e^{\varphi_y}$ are the reflection coefficients of $E^i_x$ and $E^i_y$, respectively The metasurface comprising the proposed meta-atoms ( see \textcolor{blue}{Fig. 2}) as an anisotropic-surface manifests distinct EM responses to the $x-$ and $y-$polarized incident waves, which would be evident in different $\Gamma_x$ and $\Gamma_y$. When the electric field of the incident LP wave is tilted at $45\degree$ relative to the $x-$axis, such that the phase difference of the reflected wave $\varDelta\varphi_{xy}= 90\degree$ and $|\Gamma_x|=|\Gamma_y|$,  a QWP can be produced. On the other hand, when a $y-$polarized ($x-$polarized) wave illuminates a metasurface so that the reflected wave be in $x-$axis ($y-$axis), a linear-to-linear polarization converter will be achieved.
	The scattering parameters of the proposed metasurface at different power levels in addition to the distribution of the L$-$shape and $\Gamma$-shape meta$ - $atoms determine the electromagnetic performance of the metasurface. Here, the functionality of the metasurface as a QWP at low power levels and DSWMM at high power levels is investigated  by using the commercial program, CST Microwave Studio. Periodic boundary conditions in $x-$ and $y-$directions along with Floquet ports assigned in $z-$ direction form an infinite array from the meta-atoms of \textcolor{blue}{Fig. 2(a), (b)}. At low power levels, as illustrated in \textcolor{blue}{Fig. 1}, the metasurface is illuminated by a $ 45\degree $ tilted plane wave and plays the role of a QWP; in this case,  PIN-diodes are in off-state and they are replaced  by $LC-$circuits in the full-wave simulator. \textcolor{blue}{Fig. 3(a)} shows the simulated reflection coefficients and phase delay of the QWP. At low power levels, the metasurface as the QWP presents a reflection coefficient more than $-3 dB$ and phase delay $90\degree\pm10\degree$ from 13.24 GHz to 16.38 GHz for both of $L-$ and $\Gamma-$shaped meta-atoms. This means that the metasurface converts the LP incident  wave to a CP wave in the reflection mode. 
	
	In contrast, when the nonlinear meta-atoms are exposed to high-power plane waves,  digital EM response is observable.  In fact, when the power level of $y-$polarized incident wave is increased, the PIN-diodes change their states from off-to-on. In the full-wave simulator, they are replaced with RL-circuits (see \textcolor{blue}{Fig. 2}). Therefore, the function of the metasurface is switched to an LP-to-LP polarizer.  \textcolor{blue}{Figs. 3(b)} depicts the co- and cross-polarized reflection spectra for the infinite array of L$-$shape and $\Gamma-$shape upon illuminating by $y-$polarized plane waves. In which,  $\Gamma_{xy}={E_x^r}/{E_y^i}$ and $ \Gamma_{yy}={E_y^r}/{E_y^i}$ indicate the reflection coefficients of the y-to-x and y-to-y polarizations, respectively. The cross-polarized reflection coefficient $\Gamma_{xy}$ is higher than -1.5 dB in an ultra-wide frequency band from 8.12 GHz to 19.27 GHz, while the co-polarized reflection coefficient $\Gamma_{xy}$ is lower than -10 dB in this frequency range. Hence, at high power intensities, the proposed meta-atoms can efficiently convert an LP wave into cross-polarization. As illustrated in \textcolor{blue}{Figs. 3(c)} at high power levels, there is a $ 180\degree $ phase difference between L$-$shape and $\Gamma-$shape meta$ - $atoms. Consequently, they can be considered as $"0"$ and $"1"$ binary codes. Actually, the binary phase response required for constructing the coding metasurface is acquired at the wideband frequency range. The full-wave simulation CST setup does not support the nonlinear model of the PIN-diodes, hence, the scattering parameters for both high and low power illuminations with the help of Large Signal Scattering Parameter (LSSP) solver in ADS are calculated, where the SPICE model of the PIN-diode is used (see \textcolor{blue}{Fig. 3}).
	
	\begin{figure*}[t]
		\centering
		\subfigure[][]{%
			\label{fig:23}%
			\includegraphics[height=2.02in]{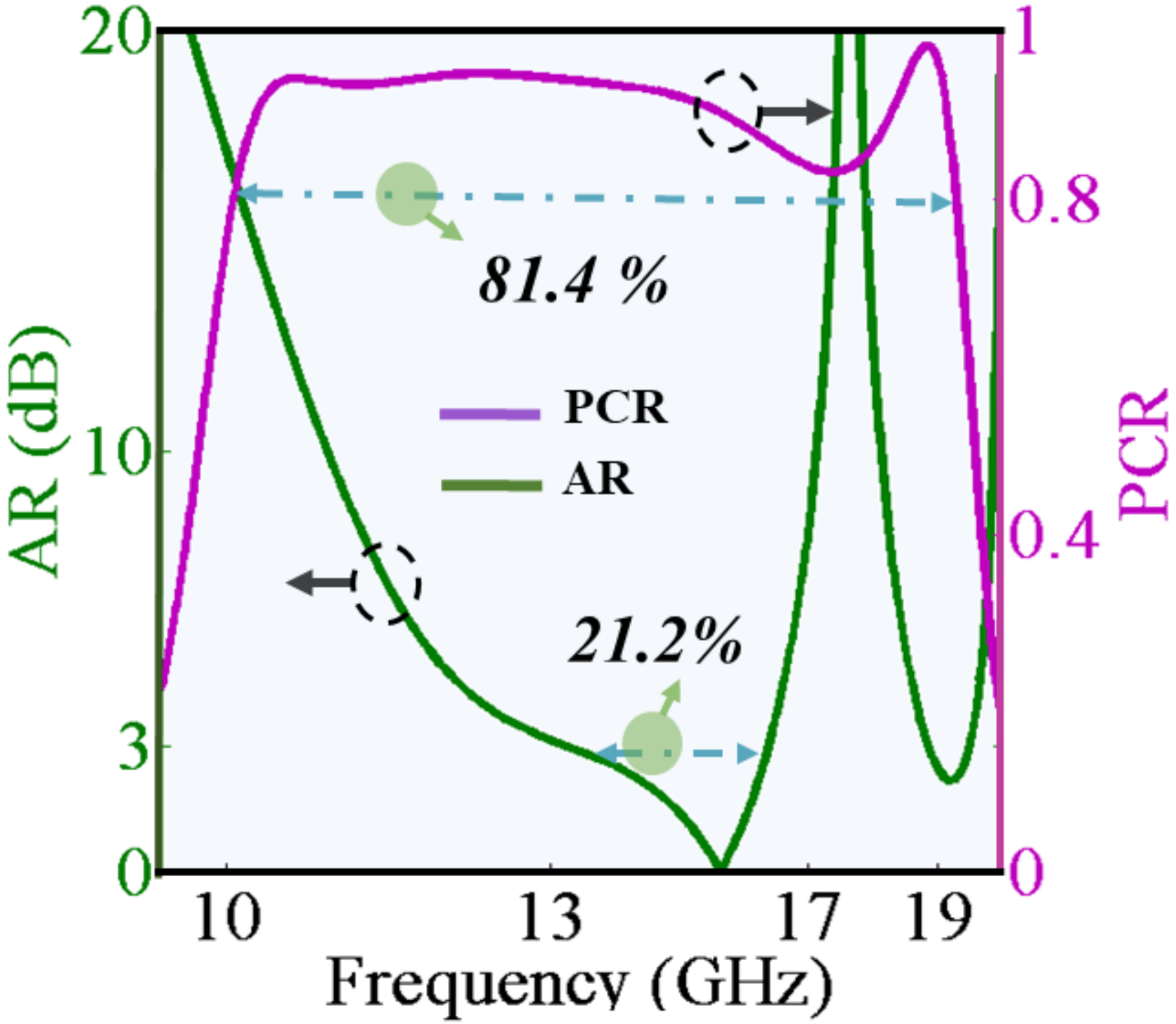}}%
		\qquad
		\subfigure[][]{%
			\label{fig:23}%
			\includegraphics[height=2.02in]{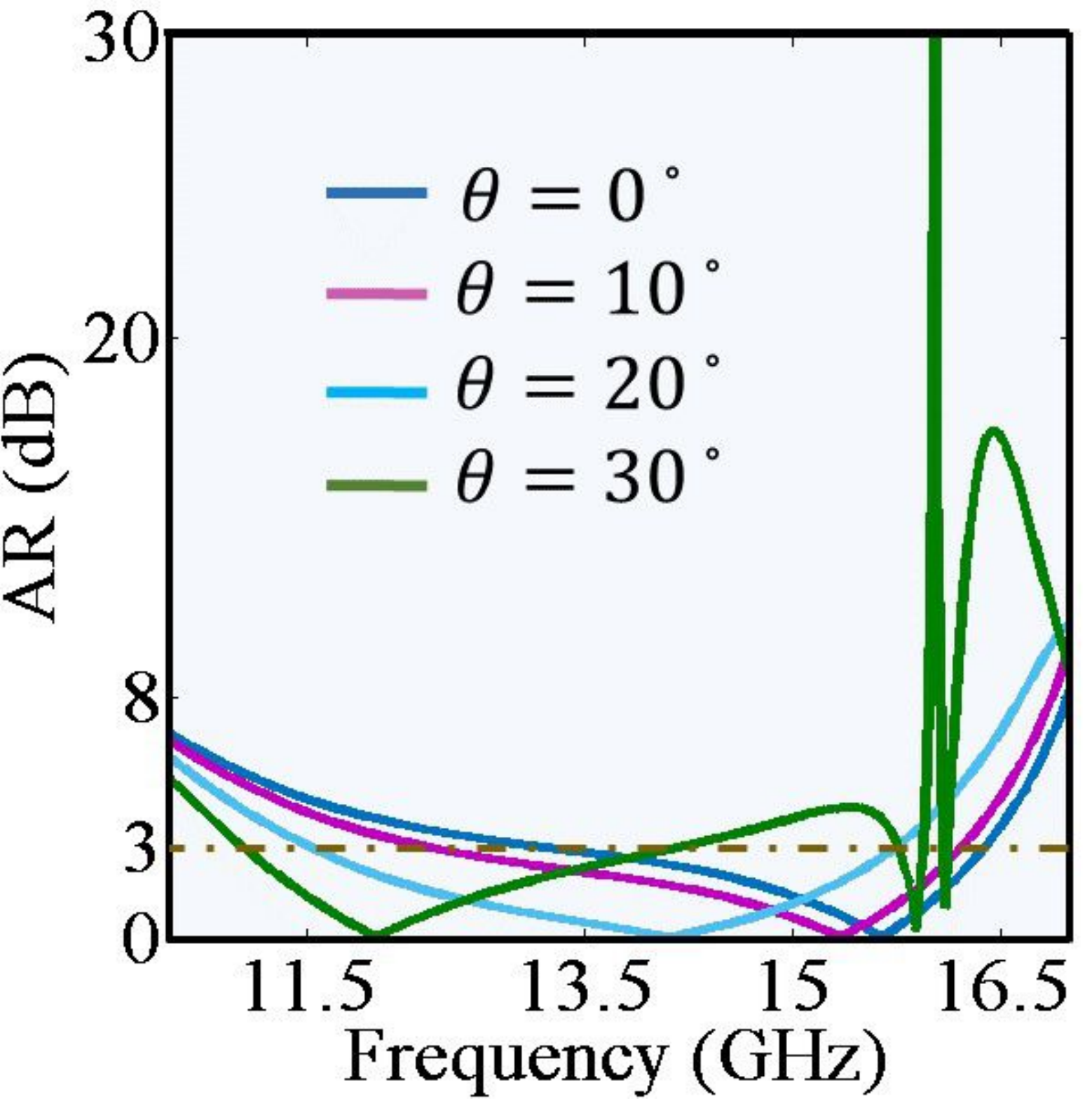}}%
		\subfigure[][]{%
			\label{fig:23}%
			\includegraphics[height=2.02in]{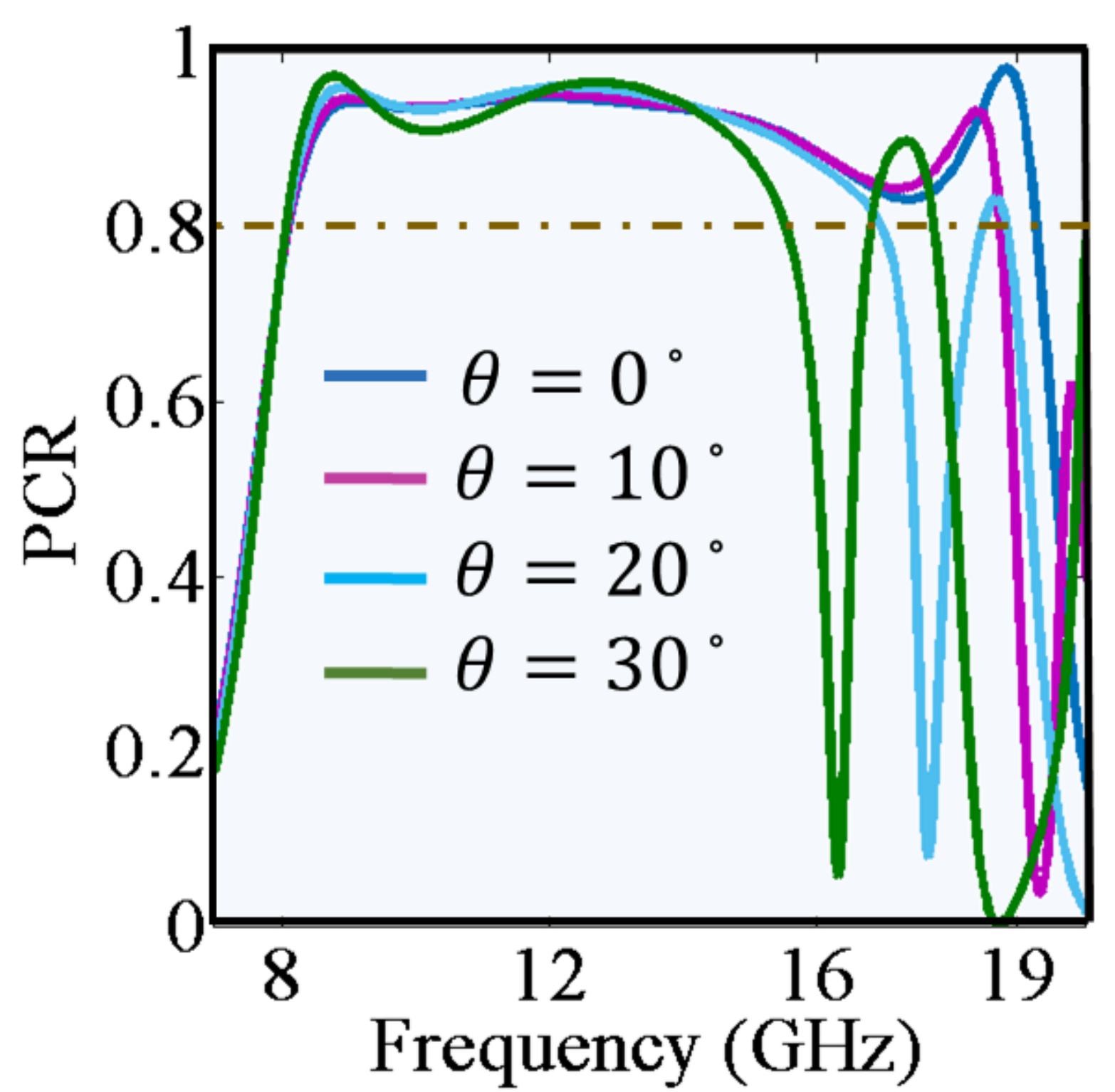}}%
		\caption{\label{fig:epsart} {(a) The calculated PCR for linear polarization conversion at high power intensities and AR for circular polarization conversion at low power intensities. The simulated (b) PCR and (c) AR at oblique incident of EM waves (equals 0\degree, 10\degree, 20\degree, and 30\degree). }}
	\end{figure*}
	
	To investigate the capacity of the proposed nonlinear anisotropic metasurface in manipulating the polarization states of reflected waves at different power levels, we calculated axial ratio (AR) for circular polarization conversion at low power levels and polarization conversion ratio (PCR) for linear polarization conversion at high power levels as follows: 
	\begin{equation}
	AR = \sqrt{\frac{\frac{1}{r}\cos^2{\alpha}+\sin{2\alpha}\cos{\Delta\varphi_{xy}}+r\sin^2{\alpha}}{\frac{1}{r}\sin^2{\alpha}-\sin{2\alpha}\cos{\Delta\phi_{xy}}+r\cos^2{\alpha}}}
	\end{equation}
	where
	\begin{equation}
	r = \frac{|\Gamma_x|}{|\Gamma_y|}
	\end{equation}
	and 
	\begin{equation}
	2\alpha = \tan^{-1}({\frac{2r}{1-r^2})\cos{\Delta\varphi_{xy}}}
	\end{equation}
	\begin{equation}
	PCR = \frac{\Gamma_{xy}^2}{\Gamma_{yy}^2+\Gamma_{xy}^2}
	\end{equation}
	
	\textcolor{blue}{Fig. 4(a)} shows the AR and PCR of the L$-$shape and  $\Gamma-$shape meta$ - $atoms at low power levels as a QWP and as a DSWMM at high power levels, respectively. It is clearly seen the AR better than $3$ dB over $21.2 \%$ from 13.24 GHz to 16.38 GHz and PCR better than $0.8$ over $81.4\%$ from 8.12 GHz to 19.27 GHz. In addition, we study the effect of oblique incidence wave on polarization conversion at different power levels. \textcolor{blue}{Fig. 4(b)} illustrates at low power intensities, altering the incidence angle from $\theta = 0\degree$ to $30\degree$ contributes to the shifting in the LP-to-CP bandwidth. On the other hand, at high power intensities, by increasing the incidence angle from $\theta = 0\degree$ to $30\degree$, the PCR bandwidth slightly starts to decreasing (see \textcolor{blue}{Fig. 4(c)}).   
	
	To justify the LP-to-RHCP performance of the metasurface at low power intensities, when PIN-diodes are in off mode, the electric field intensity distribution of the L$-$shape and $\Gamma-$shape unit cells along with phase at $14$ GHz has been shown in \textcolor{blue}{Figs. 5(a), (b)}. By increasing the phase, the light blue part on the top layer of the unit cells rotates counterclockwise that results in the radiation of RHCP waves. To further understand the working mechanism of the metasurface as the ultrawideband linear-to-linear polarizer at high power levels, the surface current of the meta$ - $atoms and ground planes at 8.6 GHz, 11.5 GHz, 15 GHz, and 18.5 GHz as resonance frequencies are depicted in \textcolor{blue}{Figs. 6(a)-(d)}. At 8.6 GHz and 15 GHz, the meta$ - $atoms currents and ground plane currents are anti-parallel; these loop currents have resulted in magnetic resonances. Conversely, at 11.5 and 18.5 GHz, the parallel currents flowing between meta$ - $atoms and ground plane have generated electric resonances. By optimizing the structural parameters, the multi-resonance metasurface can operate as an LP-to-LP polarization converter at high incident power levels in an ultra-wide frequency range.  
	
	\begin{figure}[t]
		\centering
		\includegraphics[height=3in]{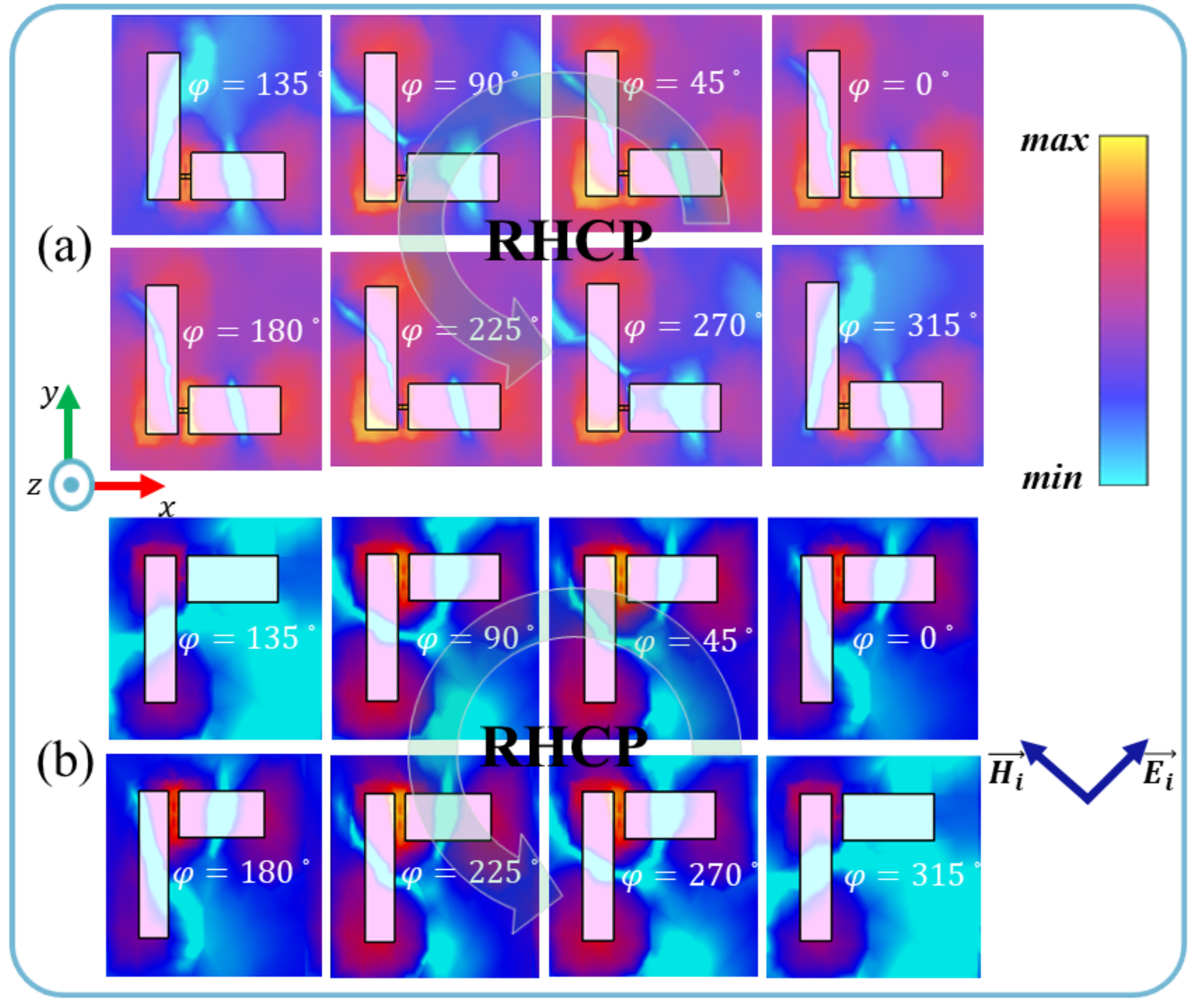}
		\caption{\label{fig:epsart} {Electric field distributions of (a) L$-$shape (b) $\Gamma-$shape meta$ - $atoms at low power intensities when PIN-diodes are in off mode at 14 GHz.}}
	\end{figure}
	\begin{figure}[t]
		\centering
		\includegraphics[height=2in]{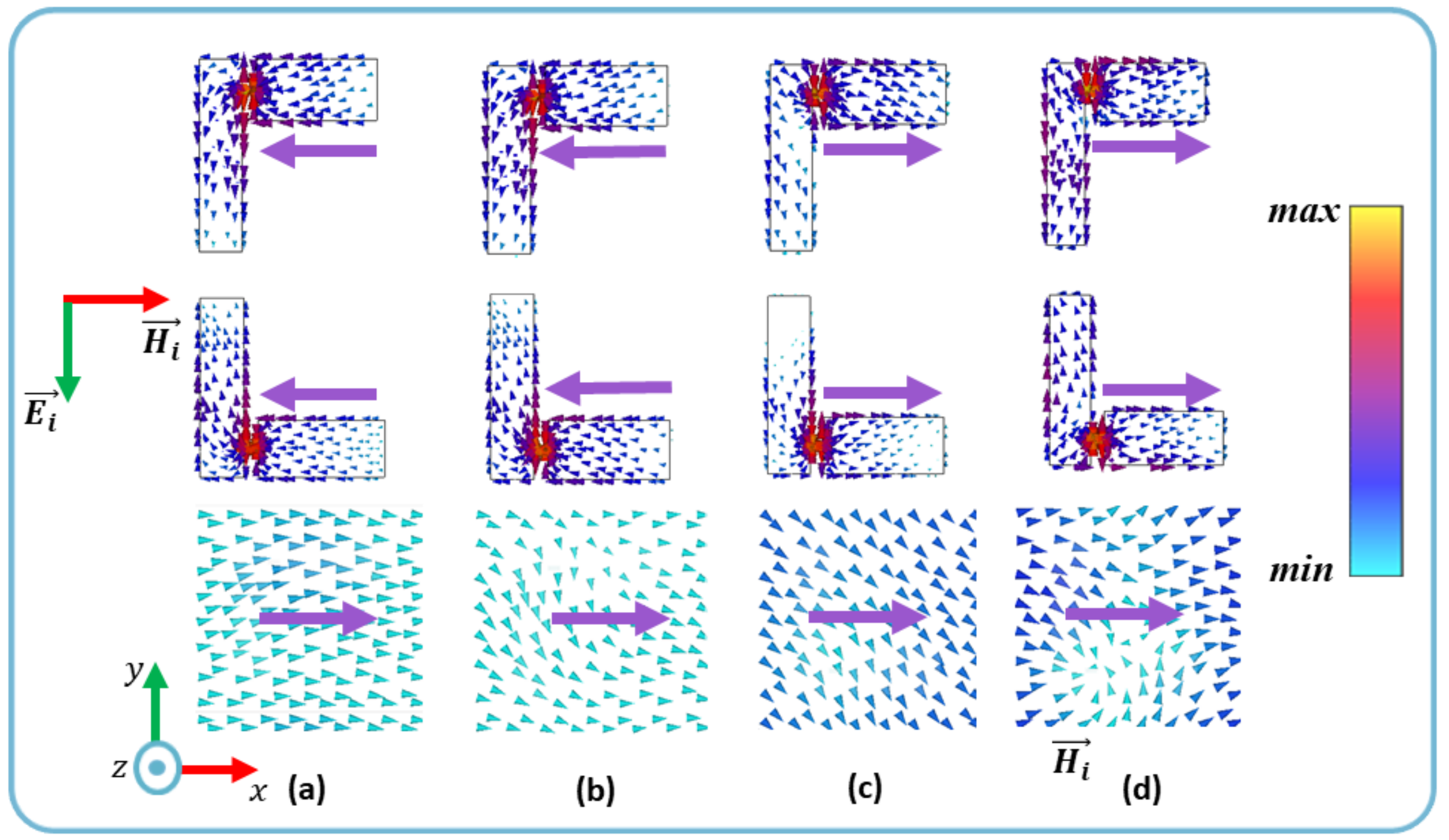}
		\caption{\label{fig:epsart} {Surface current distribution on the L$-$shape and $\Gamma-$shape meta$ - $atoms when PIN-diodes are in on mode at (a) 8.6 GHz, (b) 11.5 GHz, (c) 15 GHz, and (d) 18.5 GHz}}
	\end{figure}
	\begin{figure*}[t]
		\centering
		\subfigure[][]{%
			\label{fig:23}%
			\includegraphics[height=2.02in]{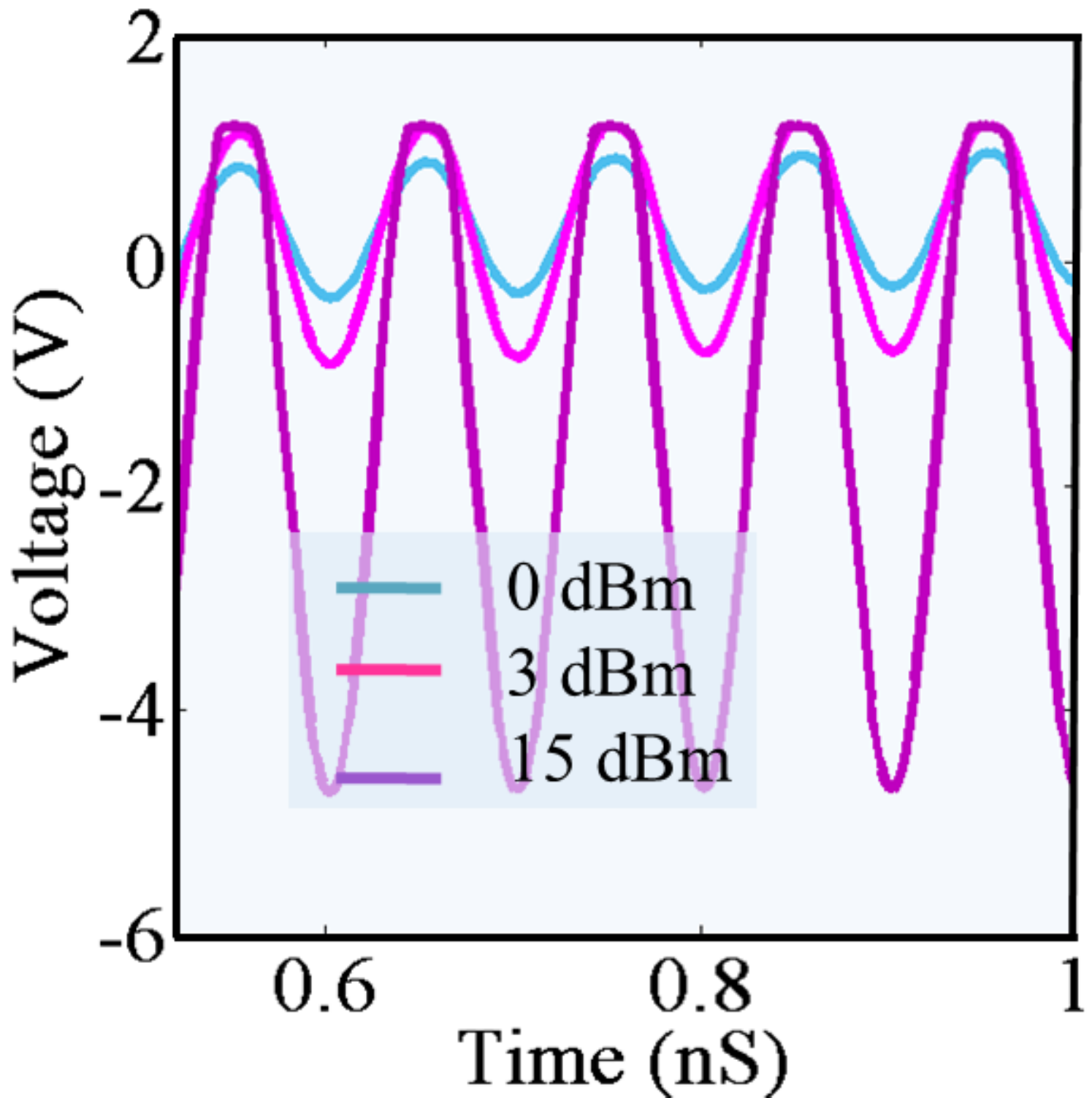}}%
		\qquad
		\subfigure[][]{%
			\label{fig:23}%
			\includegraphics[height=2.02in]{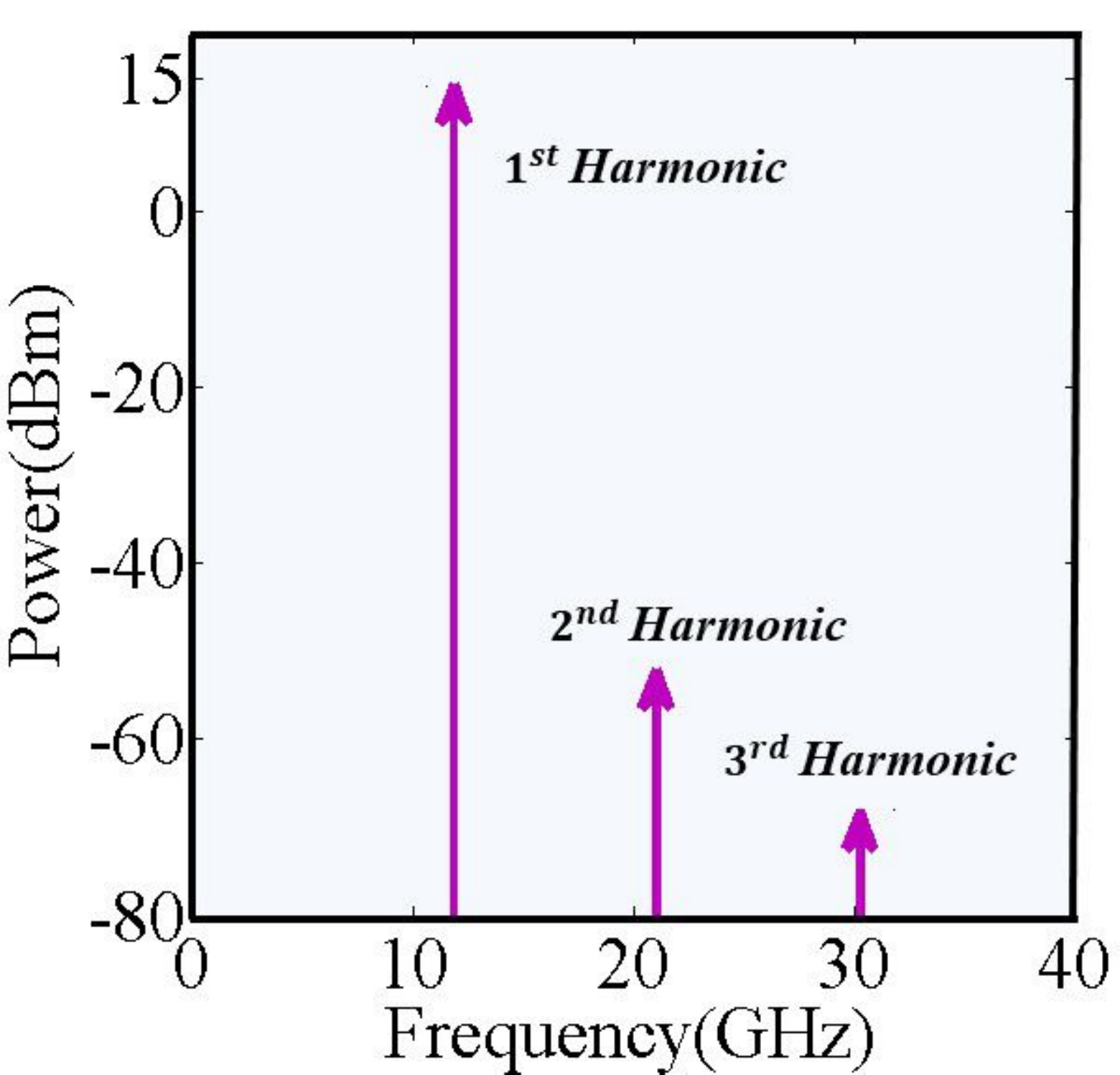}}%
		\caption{\label{fig:epsart} {(a) Induced voltage at both ends of one PIN-diode. (b) The Harmonic Balance analysis for a 15 dBm signal at 11 GHz. }}	
	\end{figure*} 
	\subsection{Nonlinear analysis}
	
	The use of PIN-diodes as the nonlinear elements may contribute to harmful effects such as harmonic generation, third-order inter-modulation and AM modulation on fundamental frequency.  Therefore, to check the validity of the structure, nonlinear analysis is essential. In this regard, the scattering parameters of the L$-$shape and $\Gamma-$shape meta$ - $atoms, in the form of an S3P file is imported to Advanced Design System (ADS)  as a circuit-based nonlinear simulator. The S3P file consists of three ports: (1), (2) the first and second Floquet ports, and (3) a discrete port instead of the PIN-diode. In the ADS medium, the discrete port is occupied with the Spice model of a Flip Chip diode with a serial number of MADP-000907-14020. In ADS as an EM-circuit simulator, using the transient solver at 10 GHz frequency and two employed probes at both ends of the PIN-diode the AC voltage of the diode at different power levels has been measured. As can be seen in \textcolor{blue}{Fig. 7(a)}, when small signals illuminate  the meta$ - $atoms, the diode voltages is a complete sinusoidal curve. By contrast, the illum$  $inated large signals generate distorted voltage wave-forms so that the positive half cycles are truncated to the diode forward voltage (1.3 V). The threshold impinged power of off-to-on switching is obtained 3 dBm for each 10 mm $ \times $ 10 mm unit cell. For the proposed metasurface that composed of 576 unit cells, the minimum impinging power to switch the state of diodes from off-to-on is calculated 30.06 dBm, therefore, the metasurface at this power as a self-biased nonlinear surface without needs to a complex DC biasing network could switch its functionality from a QWP to a DSWMM.  Again, it must be mentioned that at high frequencies for high power illuminations since the carriers injecting rate to the intrinsic layer during the positive cycles is much more than the carriers outgoing rate from the intrinsic layer during the negative cycles, despite the time variations of the induced voltage, the operational status of the diodes would be stable \cite {wu2019energy}.
	
	As mentioned before, one of the harmful effects of using nonlinear devices such as PIN-diodes may be the generation of frequency harmonics. Hence, A harmonic balance analysis using ADS software to more study the nonlinear behavior of the integrated meta$ - $atom with PIN-diode is done. In this case, one of the Floquet ports is excited with a 15 dBm signal, and the power at another port is measured. As we expected the most parts of the input power has appeared in the second Floquet port (cross-polarization) at the fundamental frequency here 11 GHz (see \textcolor{blue}{Fig. 7(b)}). As a result, the nonlinear analyzes (LSSP, Transient Response and Harmonic balance) with EM-circuit simulator validate our assumption in the modelling of PIN-diode for full-wave simulations. 
	\begin{figure*}[t]
		\centering
		\includegraphics[height=2.7in]{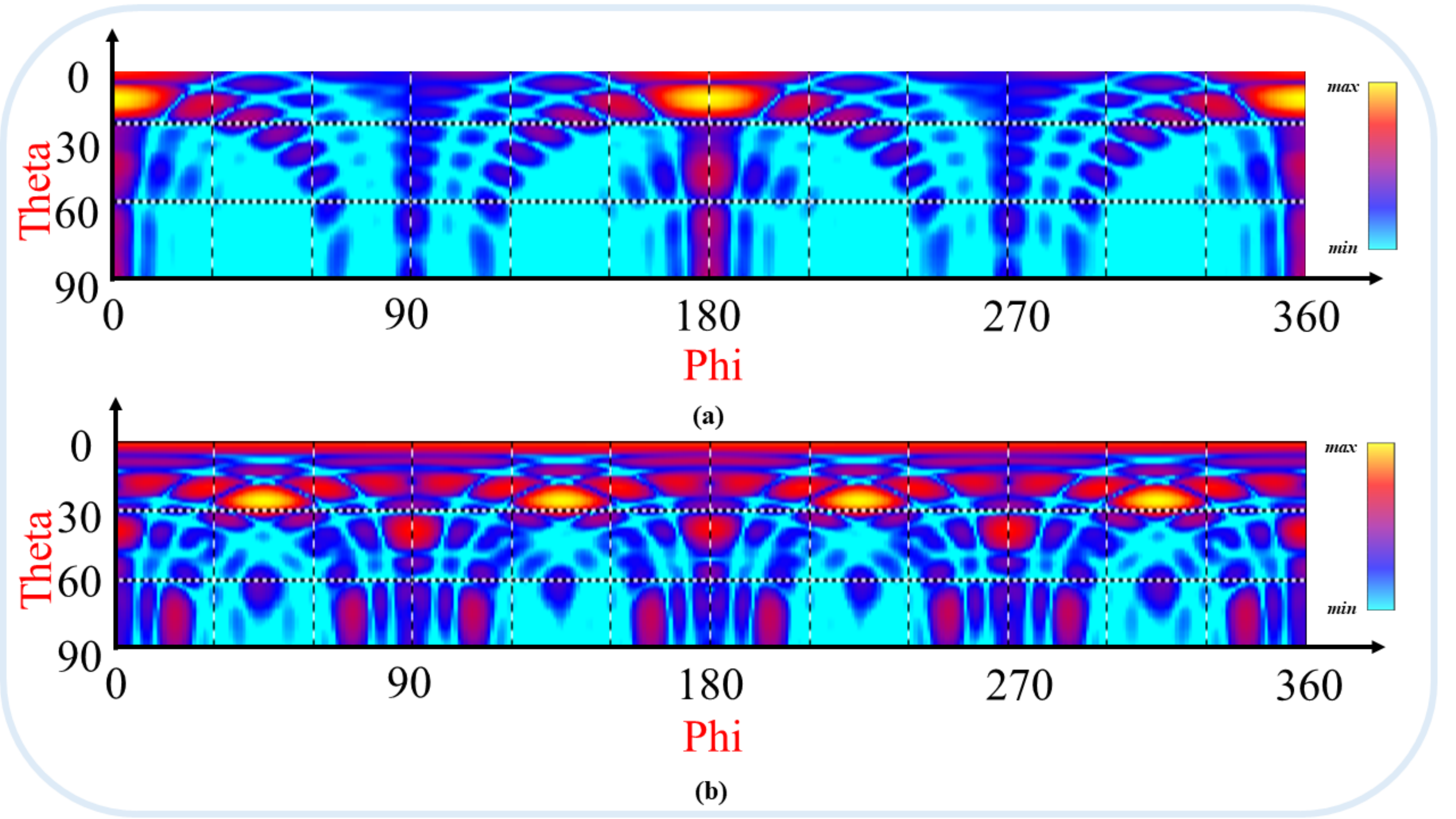}
		\caption{\label{fig:epsart} {The 2-D far-field results that show the performance of the metasurface upon illuminating by high-power incidences for (a) stripped configuration and (b) Chessboard configuration. }}
	\end{figure*}
	\section{Discussion}
	As shown in the schematic diagram of the proposed metasurface (\textcolor{blue}{Fig. 1}), at high power intensities, the distribution of the L$-$shape and $\Gamma-$shape anti-phase meta$ - $atoms as the coding particles over the metasurface spatially manipulates the incoming wave. The nonlinear metasurface as a DSWMM contains $6\times6$ binary super-cells such that each super-cell is composed of $4\times4$ L$-$shape or $\Gamma-$shape meta$ - $atoms to depreciate defective corner-related coupling effects between the neighboring super-cells. The far field pattern function of the DSWMM can be expressed by a superposition from the contributions of all super-cells. 
	\begin{widetext}
		\begin{equation}
		E^{\text{far}}(\theta ,\varphi )={f_e}(\theta ,\varphi )\sum\limits_{m = 1}^N 
		{\sum\limits_{n = 1}^N {exp} }
		\bigg( - i\big\{ {\varphi _{mn}} +kDsin\theta\big[(m - 1/2)cos\varphi  + (n - 1/2)sin\varphi \big]\big\} \bigg)
		\end{equation}
	\end{widetext}
	\begin{figure*}[t]
		\centering
		\includegraphics[height=3.7in]{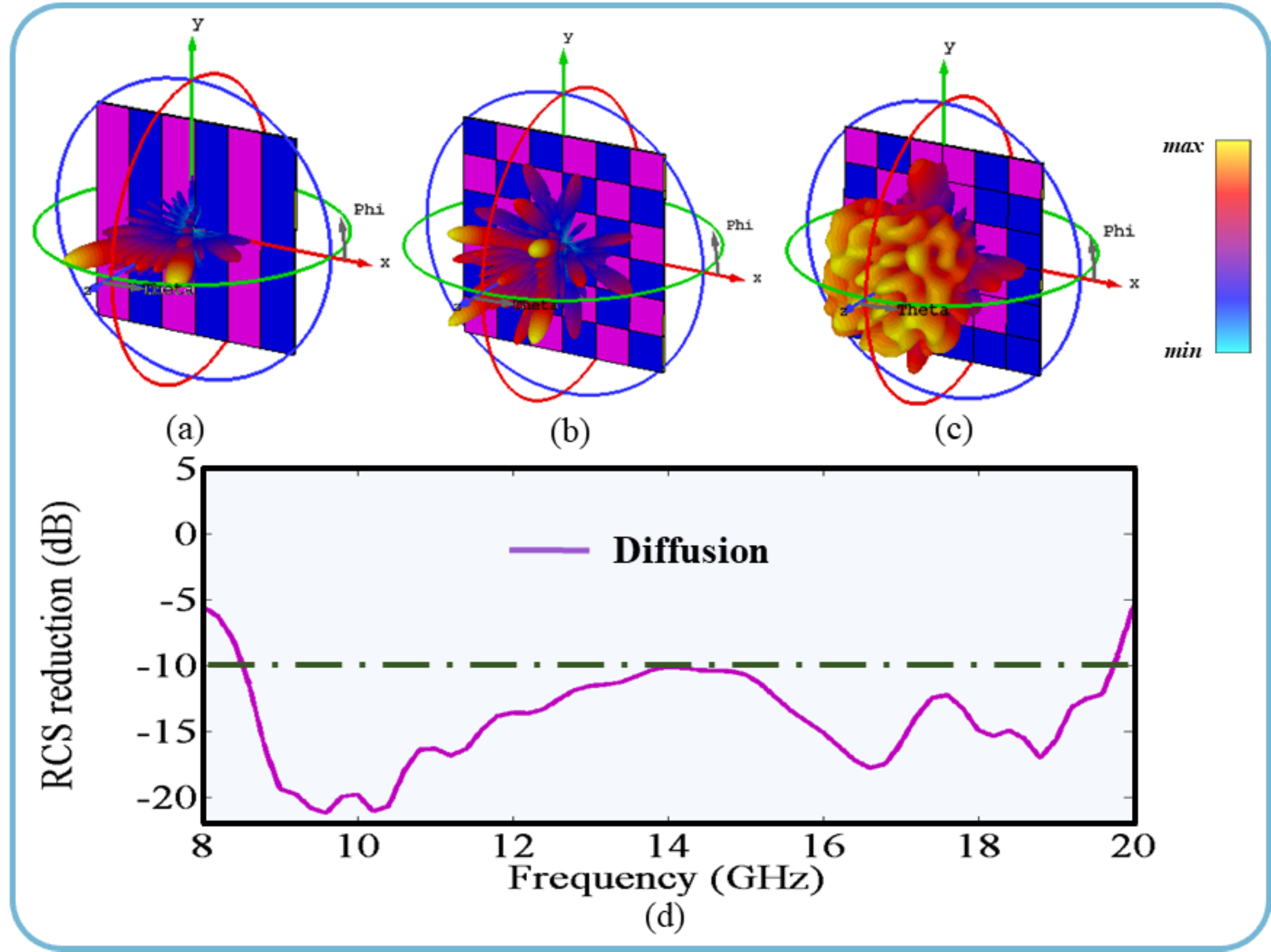}
		\caption{\label{fig:epsart} { The 3-D simulation patterns of the nonlinear metasurface as a DSWMM, which indicate its ability to manipulate reflection waves using various coding sequences under the normal incidence of EM waves at center frequency 14 GHz. (a) stripped, (b) Chessboard, and (c) the optimal diffusion configurations. (d) The ability of the randomized coding pattern at high power intensities in RCS reduction relative to the same-sized copper plate.}}
	\end{figure*}
	where, $f_e(\theta ,\varphi )$ is the pattern function of a super-cell, and $\theta$ and $\varphi$  denote to the elevation and azimuth angles, respectively. $D$ remarks the period of the digital super-cells along both x- and $y-$directions, and $k=2\pi/\lambda$ refers to the free-space wave-number. Referring to the above equation, if we consider a quasi-omnidirectional far-field pattern for all super-cells, the realized scattering patterns of different coding configurations can be expressed in the forms of two-dimensional inverse Fourier transforms of the super-cells, i.e., $E^{far}(\omega_1, \omega_2) = F^{-1} (u_{mn} )$. Here,  $\omega_1$ and $\omega_2$ remark the spatial Fourier space variables, and $u_{mn}$ is the complex status of each super-cell \cite {rajabalipanah2019ultrabroadband}.	
	
	Based on the generalized Snell's law \cite{yu2011light}, the scattering direction of the DSWMM in spherical coordinates ($\theta$, $\varphi$) can be determined using the following relations: 
	
	\begin{equation}
	{\varphi _1} =  \pm ta{n^{ - 1}}\frac{{{D _x}}}{{{D _y}}},~~ {\varphi _2} = \pi  \pm ta{n^{ - 1}}\frac{{{D _x}}}{{{D _y}}}
	\end{equation}
	\begin{equation}
	\theta  = si{n^{ - 1}}(\lambda \sqrt {\frac{1}{{D _x^2}} + \frac{1}{{D _y^2}}} )
	\end{equation}
	in which, $D_x$ and $D_y$ indicate the spatial periods of the coding sequence along the $x-$ and $y-$axis, respectively. In \textcolor{blue}{Figs. 8 (a), (b)} 
	and \textcolor{blue}{Figs. 9(a), (b)}, the normalized two-dimensional (2D) and 3D scattering pattern of the two conventional coding configurations 010101.../010101... (stripped configuration) and 010101.../101010... (chessboard configuration) at 12.6 GHz are sketched, respectively. For sequence 1, where $D_x$ is infinite and $D_y$ is 80 mm, two symmetric reflected beams with directions $(\theta_1,\phi_1)$=(17.3$^\circ$,0$^\circ$) and $(\theta_2,\phi_2)$=(17.3$^\circ$,180$^\circ$) are achieved; similarly, for chessboard configuration, four symmetric reflected beams with directions $(\theta_1,\phi_1)$=(25$^\circ$,45$^\circ$), $(\theta_2,\phi_2)$=(25$^\circ$,135$^\circ$), $(\theta_3,\phi_3)$=(25$^\circ$,225$^\circ$), and $(\theta_4,\phi_4)$=(25$^\circ$,315$^\circ$) are obtained. The numerical results in \textcolor{blue}{Figs. 8 (a), (b)} are obtained with the help of CST Microwave Studio, which is consistent with  theoretical predictions. The use of randomness distribution of the coding particles gives rise to lower scattering signatures over a broad frequency range from X-band to Ku-band. Hence, we use an entropy-based method to reach the efficient diffusion-like scattering pattern for smart camouflage surface applications. Entropy as a key gauge to describe the information of different types of coding configuration Can be accepted \cite{golshani2009some}. Surely, the more average entropy measure, as well as diffusion level, results in the lower probability of estimation of scattering patterns. In \cite{momeni2018information}, the authors have validated the ability of the 2D quadratic Renyi entropy to reach the best diffusion-like coding configuration. According to the high ability of the Renyi entropy in measuring Gaussianity in the 2D scattering pattern, it could be utilized to interpret different coding configurations. Herein, Binary Bat Algorithm (BBA) as a fast, simple, and efficient optimization method in comparison to the conventional optimization methods to reach the best diffusion phase-encoded pattern has been used that its relation can be expressed as:
	
	\begin{equation}
	H_{Re} = \text{max} \bigg \{ {\frac{-1}{2} \text{log} \bigg[\sum\limits_{i = 1}^{128} {\sum\limits_{i_R = 1}^{128} {P_{ii_R}^2}} + \sum\limits_{i = 1}^{128} {\sum\limits_{i_R = 1}^{128} {P_{ii_D}^2}}}\bigg] \bigg \}
	\end{equation}
	here, $P_{ii_R}$ and $P_{ii_D}$ remark the $ii_R$ and $ii_D$ indexes in $\text{PDF}_{JR}$ and $\text{PDF}_{JD}$ diagrams, respectively. These two $\text{PDF}$ diagrams based on the statuses of each neighbouring pair of the 2D-IFFT image matrix elements in the vertical and horizontal directions can be determined.  
	The Renyi entropy for different distribution of L$-$shape $ \Gamma- $and shape meta$ - $atoms as the anti-phase coding particles is calculated and shown in \textcolor{blue}{Table. 1}. In addition, with the help of Binary Bat Algorithm (BBA) the optimal diffusion coding configuration to reach the maximum entropy, as well as the maximum diffusion among all potential solutions, is achieved. In this case, the quantity of the $H_{Re}$ is calculated as high as 1.80. As can be seen in \textcolor{blue}{Figs. 9(c)}, the high power incident waves are scattered along with the numerous random directions by the entropy-based randomized coding metasurface in such a way no main scattered lobes exist. As an application, this nonlinear metasurface can reduce both  bi-static and mono-static RCS (see \textcolor{blue}{Figs. 9(d)}) for high power incident waves. The numerical simulations clarify the significance of this work at the microwave band among all of the nonlinear metasurfaces that recently are introduced. Unlike previous works,  the metasurface according to the power level of the incoming waves controls and manipulates them over 21.2\% (low power signals) and 81.4\% (high power signals) bandwidths.  
	\begin{table}
		\caption{\label{tab:table1}%
			The computed Renyi entropy for different distribution of coding particles. 
		}
		\begin{ruledtabular}
			\begin{tabular}{lcdr}
				\textrm{Coding configuration}&
				\textrm{Renyi entropy($ H_{Re} $)}
				\\
				\colrule
				PEC reflector & 0.143 \\
				Stripped configuration & 0.318\\
				Chessboard configuration & 0.452\\
				Optimal Diffusion configuration & 1.80\\
			\end{tabular}
		\end{ruledtabular}
	\end{table}
	\section{Conclusion}
	We proposed a passive wideband self-biased nonlinear reflective metasurface at microwave frequencies to spatially control electromagnetic waves. The designed metasurface is composed of L$ - $shape and $\Gamma -$ shape meta$ - $atoms that loaded with self-biased PIN-diode elements. The EM functionality of the passive structure is determined by the level of incident power so that at low power intensities, the nonlinear metasurface as an anisotropic surface acts as a QWP, changes the linear polarization of the incoming wave to the circular polarization.
	On the other hand, the multi-resonant characteristic of the nonlinear metasurface at high power intensities causes the linear polarization of the incident wave converts to cross-polarization. Moreover, using the anti-phase meta$-$atoms with same magnitudes results to a digital metasurface. Then, an entropy-based binary coding pattern and Binary Bat Optimization Algorithm are mixed to obtain a diffusion-like far-field pattern over a broad frequency range. Nonlinear circuit-based analyzes in addition to the full-wave simulations verified the powerful wave control ability of the nonlinear metasurface.

	\nocite{*}
	
	\bibliography{apssamp}
	
\end{document}